# On-chip Light Trapping in Bilayer Moiré Photonic Crystal Slabs

Haoning Tang[1⊥*], Xueqi Ni[1⊥], Fan Du[1], Vishantak Srikrishna[2], Eric Mazur[1*]

[1] School of Engineering and Applied Sciences, Harvard University, Cambridge, MA 02138, USA

[2] Department of Physics, University of California, Berkeley, California 94720, USA

[⊥] Authors contributed equally to this work

[*] Email: mazur@seas.harvard.edu, hat431@g.harvard.edu

**Abstract**

The optical response of bilayer moiré photonic structures can be precisely controlled by varying the lattice geometry. Bilayer moiré photonic crystal slabs exhibit flat bands in the optical band structure, where the optical modes have zero group velocity. They also give rise to momentum-independent light-trapping of Bloch waves in both transverse and vertical directions, leading to high quality-factors ($Q = 10^9$) and small mode volumes ($V = 0.8\ \lambda^2$). The large $Q$ and small $V$ lead to a large Purcell enhancement ($F_P = 300$), providing opportunities for low-threshold lasing, enhancement of optical nonlinearities, and quantum information processing.

**Introduction**

Over the past decade several approaches have been developed for light trapping in on-chip photonic devices to enhance spontaneous emission in cavity quantum electrodynamics and nonlinear optics [1-5]. Enhancing spontaneous emission necessitates a high field intensity, which requires small mode volume *V* as well as a large *Q* factor [6, 7]. Photonic crystal cavities, using defects and photonic crystal band gaps, forbid in-plane propagating waves, giving rise to highly localized modes (Figure 1a and b) [8-15]. However, photonic crystal cavities introduce a trade-off between the *Q* factor and the mode volume in the cavity — the decrease in transverse mode volume *V* increases radiative losses, causing a corresponding reduction in *Q* factor [8]. An alternative approach is to use photonic crystal slabs without defects and design optical bound-in-continuum (BIC) states that do not couple to external radiation fields (Fig. 1c-d) [16, 17]. Even though the *Q* factors of BIC photonic crystal slabs can be infinite in theory, fabrication disorder typically limits their values to $10^4$ to $10^6$ (Fig. 1g) [18-22]. In addition, they only localize modes in the vertical (thickness) direction, leaving the transverse modes delocalized across the slab, providing a lower bound on the mode volume. Furthermore, the *Q* factor of BIC photonic crystal slabs is dependent on the momentum of the electromagnetic wave, which limits their application in omnidirectional devices. It is, therefore, of interest to explore a photonic device that concentrates light in both the vertical and transverse direction, independent of the wave momentum.

In this paper we show that bilayer moiré photonic crystal slabs provide both small mode volume and high *Q* factors, offering a new, on-chip light-trapping approach. As the layers are twisted relative to each other, the moiré photonic crystals display periodicities at two different scales: periodicity of the photonic lattice and that of the superlattice[23-26]. Therefore, the underlying oscillation of the electromagnetic field is controlled by the photonic lattice periodicity, but the overall profile is modulated by the moiré superlattice periodicity. As a result, the light can be confined in certain superlattice sites (Fig. 1e-f) [27]. This confinement coincides with the appearance in the bilayer moiré photonic crystal band structure of moiré bands that have a high local density of states (LDOS) at AA (most aligned) stacking sites. At certain twist angles, the moiré bands become flat, and the group velocity of the optical modes drops to zero, giving rise to light-trapping in the transverse direction. Because moiré band modes have a high $Q$, the light is also confined in the vertical direction. The vertical and lateral confinement provides twist-angle-tunable, momentum-free trapping of Bloch waves, giving rise to high $Q$ factors and small mode volume $V$. This contrasts with most photonic crystal cavities that have localized non-Bloch modes with small $V$ and relatively low $Q$ factor or bound-states-in-continuum photonic crystals with high $Q$ factors but large $V$. We numerically demonstrate momentum-independent $Q$ factors as high as $10^9$ and momentum-independent mode volumes $V$ on the order of $0.8\ \lambda^2$ giving rise to a Purcell factor of 300. Bilayer moiré photonic crystals, therefore, are promising devices for on-chip light trapping and has potential applications in cavity quantum electrodynamics and nonlinear optics (Fig. 1g) [28, 29].

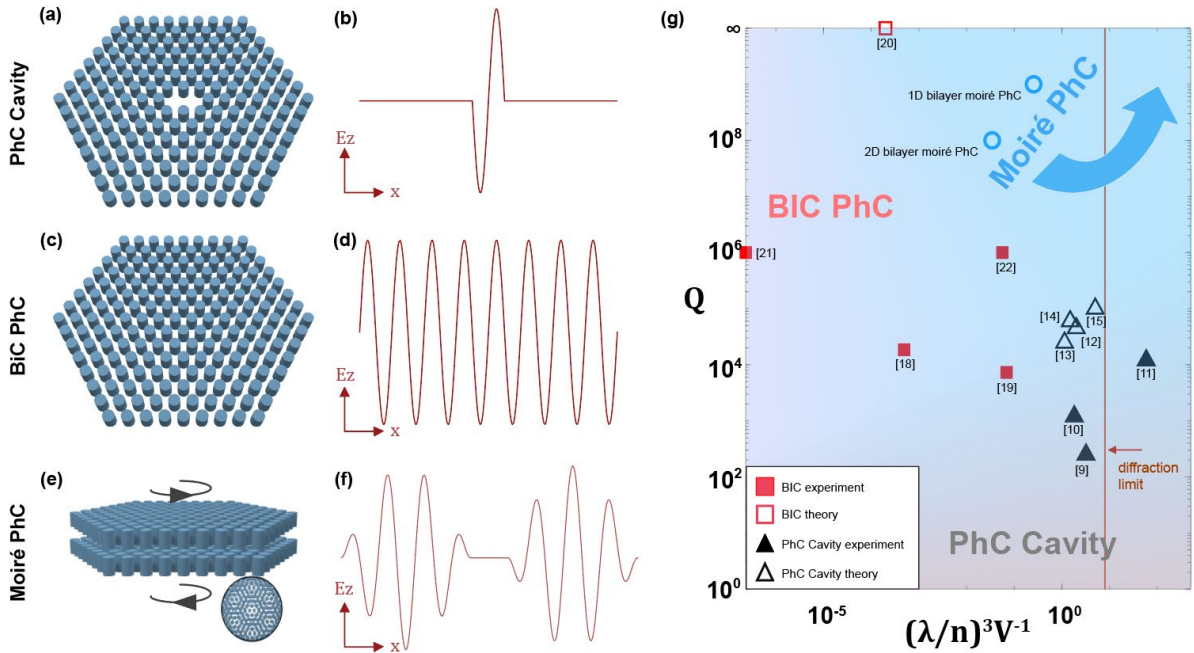

**Figure 1** Photonic crystal structures and their electromagnetic modes. (a) (b) Defect photonic crystal cavity and its in-plane light confinement; (c) (d) High $Q$ 2D photonic crystal slabs with guided Bloch modes, such as the bound-states-in-continuum photonic crystals; (e) (f) Twisted bilayer moiré photonic crystal. The amplitude of the electromagnetic field is modulated by the moiré superlattice. (g) Comparison of $Q$ factor and mode volume for photonic crystal cavities, BIC photonic crystals, and moiré photonic crystals (details in supplementary material).

**Band structure and $Q$ factors**

We started by determining band structure and $Q$ factor for both 1D and 2D moiré bilayer photonic crystal slabs (Figure 2). The 1D moiré bilayer slabs have periodicity along the in-plane axis and finite height along the out-of-plane axis (Figure 2a), while the 2D moiré slabs have periodicity along two in-plane axes and finite height along the out-of-plane axis (Figure 2d). The 1D system consists of two 180-nm thick grating layers with the same filling fraction $\kappa = 0.8$ and different lattice constants $a_1$ and $a_2$, separated by an airgap of subwavelength thickness $d = 36$ nm. The lattice constants satisfy the commensurate condition $a_1(N + 1) = a_2 N$ where $N$ is an integer number (Fig. 2a). The different lattice constants produce a moiré pattern with a macroscopic periodicity of distinct AA (aligned) and AB/BA (misaligned) stacking regions that grow in size as $N$ increases [30]. The moiré superlattice has a corresponding Brillouin zone that is smaller than that of the single layers. As *N* is increased, the two isolated bands that appear in the bandgap become increasingly flat (Fig. 2c). At the magic superlattice value $N = 19$, the bandwidth of the two bands reaches their lowest value, giving rise to the highest density-of-states at those frequencies (Fig. 2c). The *Q* factor of the corresponding flat-band modes for the 1D moiré bilayer photonic crystal slabs is $10^9$ across the entire momentum space.

The 2D moiré bilayer photonic crystal slabs consists of two identical honeycomb lattices that are twisted relative to each other (Fig. 2d) and separated by a polymethyl methacrylate (PMMA) coupling layer. Each honeycomb lattice is a 220-nm thick crystalline silicon membrane ($n_{Si} = 3.48$) with $C_{6v}$ symmetry-protected triangular air holes. The triangular holes have a side length $b = 279$ nm and a unit cell pitch $a = 478$ nm. The coupling layer has a thickness $d = 250$ nm and a refractive index $n_{PMMA} = 1.48$. The relative twisting of the photonic crystal produces a moiré pattern with distinct AA (aligned) and AB/BA (misaligned) stacking regions that grows in size as the twist angle decreases. In reciprocal space the twisting causes two sets of Dirac cones to intersect and hybridize with each other [31]. It also causes the guided resonances in the two layers to couple through their evanescent fields. As the twist angle gets smaller, the Dirac cone bands get closer to each other and hybridize into increasingly narrow bands (Fig. 2e). We previously showed that at the 'magic' twist angle $\theta = 1.89°$, the bandwidth of the flat band at 190 THz is reduced to just 0.217 THz and the group velocity becomes zero [32, 33]. As Figure 2e shows, the *Q* factor for the corresponding flat-band modes is $10^8$. The moiré band structures and *Q* factors are calculated using finite-element method. We note, however, that the precision of the simulation mesh we used is limited by computational resources, and because the *Q* factor increases as the mesh is narrowed, the *Q* factor we report here is likely to be a lower bound.

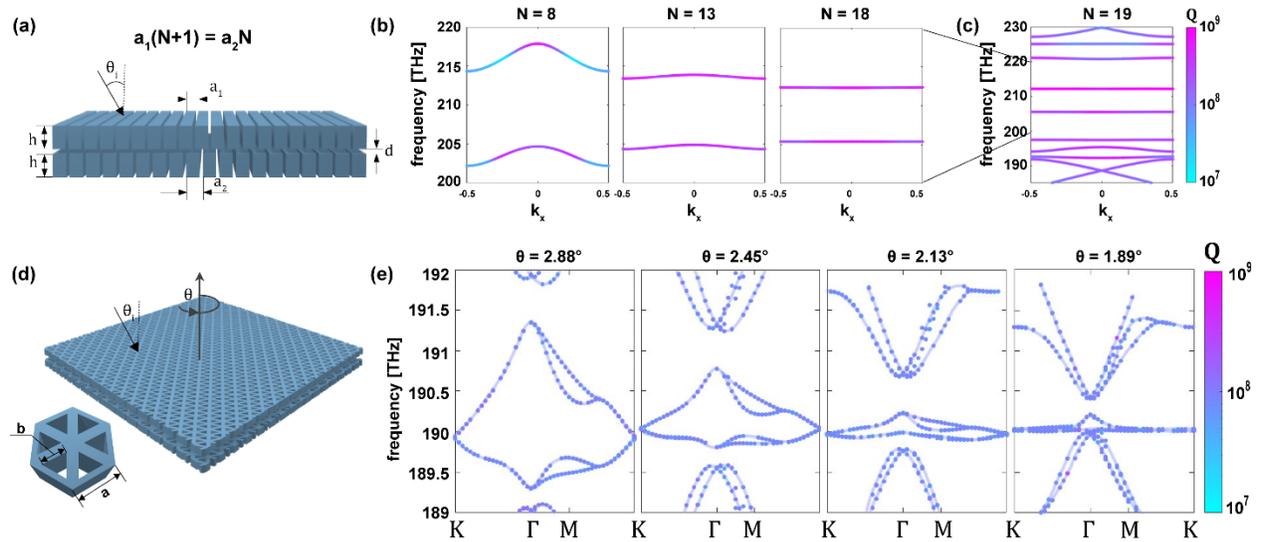

**Figure 2** Band structure for 1D and 2D moiré bilayer photonic crystal slabs. (a) The geometry of 1D moiré gratings under commensurate conditions; (b) Band structure and its evolution with respect to the number of arrays in a superlattice, *N*. Color bar indicates *Q* factor; (c) Zoomed-out band diagram showing the two optimized flat bands at the center of the vertical scale; (d) Geometry of 2D twisted bilayer photonic crystals. (e) Band structure evolution as a function of twisted angle $\theta$, showing the hybridization of Dirac cones and appearance of flat-band at around 190 THz. The color bar also indicates the *Q* factors.

**Optical response and mode volume**

Next, we examine the optical response of the 1D and 2D bilayer moiré photonic crystal slabs in the flat-band regime to an incident plane wave, while varying the angle of incidence relative to the surface normal. Figure 3 shows the transmission of the incident wave, the mode profile in the bilayer system, and the corresponding mode volume. The two resonances in the transmission curves for the 1D moiré photonic crystals (Fig. 3a) correspond to the two flat bands in Fig. 2c. The fact that these resonances are the same across incident angles from 5° to 30° demonstrates the flatness of the bands across momentum space. Figure 3b shows the mode profiles at each of the two flat band wavelengths of 1412 nm (top) and 1444 nm (bottom); the corresponding frequencies are 212.5 THz and 207.8 THz, respectively. Both mode profiles are strongly confined at AA sites. At the higher frequency, the mode profile penetrates through the coupling layer across the two photonic crystal layers (top) and at the lower frequency, the mode profile is concentrated in the photonic crystal slabs (bottom). At the flat-band frequencies, the mode profiles do not vary as the angle of incidence is varied. The mode profiles allow us to obtain the mode volume (details in supplementary material). As shown in Figure 3c, the mode volume decreases by a factor of 10 at the flat-band frequencies and is insensitive to the incident angle.

In the 2D bilayer moiré photonic crystal slabs, we do not observe any far-field optical resonances in the plane-wave transmission spectrum (Fig. 3d). The absence of resonances can be attributed to the much greater number of moiré wavevectors in the 2D system, causing the resonances to be washed out over the transmission spectrum [34, 35]. However, at flat-band frequency of 190.2 THz, the mode profile is again strongly confined at AA sites (Fig. 3e). Figure 3f shows how the mode volume again sharply decreases by a factor of 10 at the flat-band frequency and is insensitive to the incident angle (Fig. 3f). Therefore, both 1D and 2D bilayer moiré photonic crystals slabs provide momentum-independent three-dimensional light-trapping at flat band frequencies.

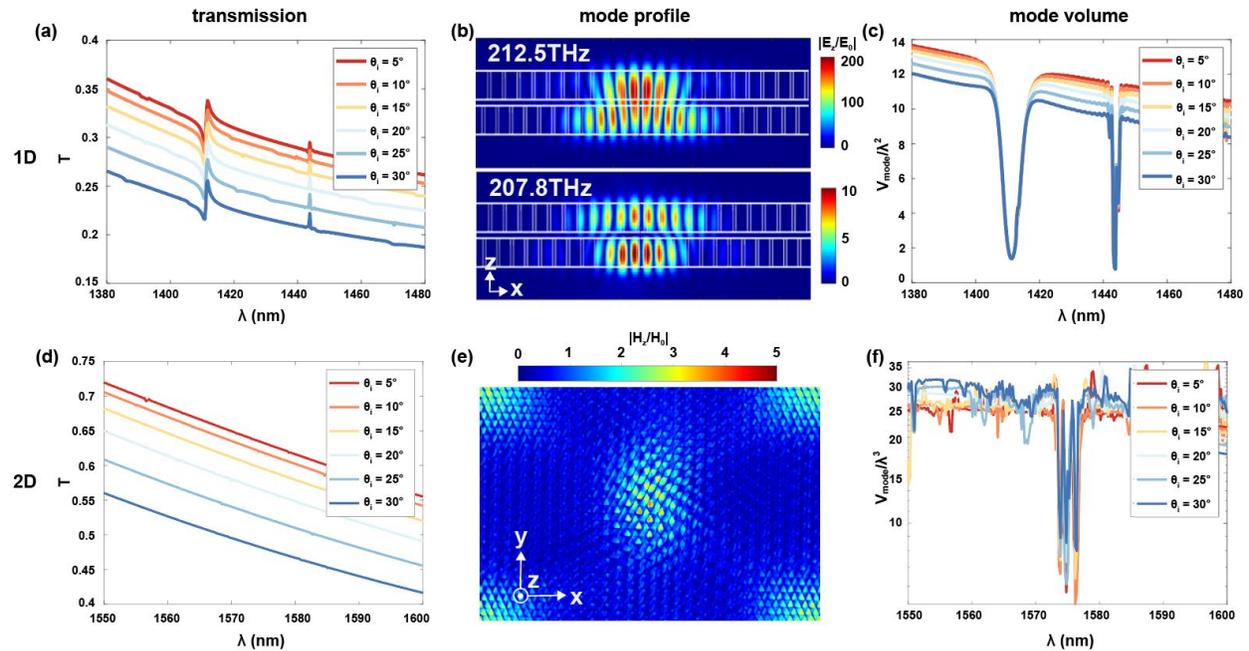

**Figure 3** Plane-wave optical responses. (a) Plane-wave incident-angle-resolved transmission spectra of the 1D moiré gratings with $N = 19$; (b) TM mode profiles at two flat band frequencies, which have been normalized to the electric field of the source, $\mathbf{E}_0$; (c) Mode volume spectra with different incident angles at $N = 19$ (see supplementary material for calculation details); (d) Incident angle-resolved transmission spectra of the 2D moiré slabs with dark resonance. (e) TE mode profile at flat-band frequency, showing light trapping at AA stacking region. The mode amplitude is normalized to the magnetic field of the source,

$\mathbf{H}_0$; (See details in supplementary) (f) Incident angle-resolved mode volume of the 2D twisted bilayer photonic crystals with twist angle $\theta = 2.13°$.

**Local density of states and Purcell factor**

Both 1D and 2D bilayer moiré photonic crystals slabs also lead to a large local density of states (LDOS) and Purcell factors. We obtained the LDOS from the imaginary part of the Green's function by sweeping the position of a single dipole point source across the structure (see supplemental materials) [36, 37]. The dyadic Green's function $\mathbf{G}(\mathbf{r}, \mathbf{r}_0)$ is defined by the electric field at the point $\mathbf{r}$, generated by a point source at point $\mathbf{r}_0$ with dipole moment $\boldsymbol{\mu}$,[38]

$$\mathbf{E}(\mathbf{r}) = \frac{\omega^2}{\varepsilon_0 \varepsilon_r c^2} \mathbf{G}(\mathbf{r}, \mathbf{r}_0) \cdot \boldsymbol{\mu}. \tag{1}$$

$\mathbf{G}$ is a symmetric $3 \times 3$ matrix and each component of $\mathbf{G}$ can be obtained from the corresponding dipole orientation and electric field component. In both 1D and 2D photonic structures, we only consider TM-like modes or TE-like modes, which can be represented by $E_z$ or $H_z$ in the $z$-direction. Consequently, we can limit our discussion to the Green's function element $G_{zz}$, which can be calculated from a dipole oriented along the $z$-direction as

$$G_{zz} = \frac{\varepsilon_0 \varepsilon_r c^2 \omega^2}{\mu \omega^2} E_z. \tag{2}$$

The corresponding local density of states can then be obtained from the imaginary part of the dyadic Green's function,

$$\rho_z(\mathbf{r}_0, \omega) = \frac{6\omega}{\pi c^2} \text{Im}\{G_{zz}(\mathbf{r}_0, \mathbf{r}_0; \omega)\}. \tag{3}$$

Figure 4 shows the calculated LDOS and Purcell factor for the 1D and 2D systems at the flat-band frequencies. For both systems, the LDOS in the AA-stacked regions is enhanced by a factor of almost three orders of magnitude over that in the AB-stacked regions (Fig. 4a and b). For the 1D system, the transverse mode profile is narrower at the higher flat-band frequency and so the higher flat-band frequency has a higher maximum LDOS than the lower flat-band frequency (Fig. 4a).

The LDOS enhancement leads to a Purcell enhancement — an increase in single-atom decay rate relative to free space [39]. The ratio of the single-atom decay rate in the material to the decay rate in vacuum, called the Purcell factor, is given by [40, 41]

$$F_P = \frac{\text{Im}[\mathbf{G}(\mathbf{r}_0, \mathbf{r}_0; \omega)]}{\text{Im}[\mathbf{G}_0(\mathbf{r}_0, \mathbf{r}_0; \omega)]}, \tag{4}$$

where $\mathbf{G}(\mathbf{r}_0, \mathbf{r}_0; \omega)$ is the dyadic Green's function of a point source in the photonic crystal slab, and $\mathbf{G}_0(\mathbf{r}_0, \mathbf{r}_0; \omega)$ is the dyadic Green's function of point source in free space. Figures 4c and d show that Purcell factors for both 1D and 2D moiré photonic crystal slabs increase by more than two orders of magnitude at the flat-band frequencies. For the 1D system, we have $F_P = 25$ at $f = 207.6$ THz and $F_P = 285$ at $f = 216.2$ THz; for the 2D system, we have $F_P = 218$ at $f = 190.3$ THz and $F_P = 173$ at $f = 190.4$ THz. The spatial enhancement of the LDOS and large Purcell factor at the flat band frequencies facilitate the control of spontaneous emission rate.

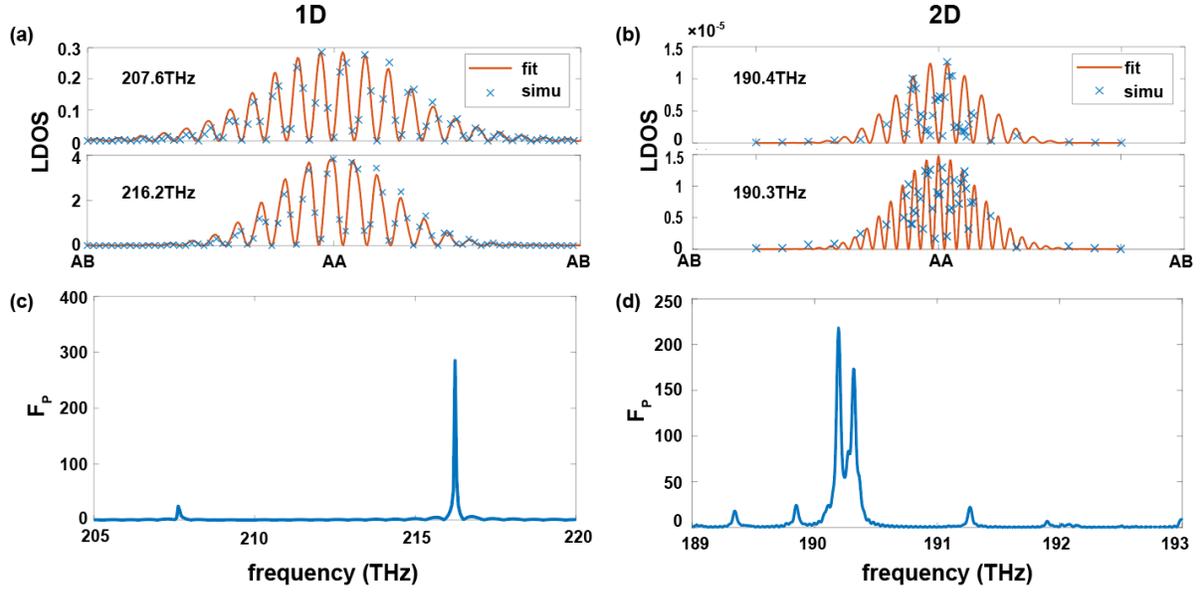

**Figure 4** (a) Local density of states in 1D moiré photonic crystal ($N = 19$) at two flat bands and (c) The Purcell factor at different frequencies. (b) Local density of states in 2D twisted bilayer moiré photonic crystal ($\theta = 2.13°$) at two flat bands and (d) the Purcell factor at different frequencies. The simulation result is fitted by the LDOS fitting function (see supplementary).

**Conclusion**

We showed how bilayer moiré photonic crystal slabs confine light in both transverse and vertical directions, enhancing the LDOS and Purcell factor. Unlike traditional photonic crystal structures, such as defect photonic crystal cavities and BIC photonic crystals which give rise to a trade-off between the *Q* factor and the mode volume *V*, bilayer moiré photonic crystals allow us to explore a new regime of high *Q* and small *V*. This regime results in a spatially enhanced LDOS and an increased Purcell factor. Given the tunability of the moiré lattice, the increased Purcell factor can be used to selectively enhance the spontaneous emission of specific emitters. In addition to the spatial confinement, the flat bands also cause the group velocity of the optical modes to drop to zero, opening the door to slow-light effects [42, 43]. The spatial confinement and zero group velocity of the moiré structures permit localizing light-matter interactions for a range of applications, including low-threshold lasing, single-photon sources, quantum electrodynamics, photonic circuits, and quantum information processing.

The supplementary material outlines in greater detail the mathematical and simulation methods we used to obtain the band structures, *Q* factors, optical responses, LDOS, and Purcell factors.

Author Contributions

[⊥]H.T. and X.N. contributed equally to this work.

Several people contributed to the work described in this paper. H.T. conceived the basic idea for this work. H.T., X.N., and F.D. carried out the simulations and analyzed the results. E.M. supervised the research and the development of the manuscript. X.N. prepared the first draft of the figures. H.T. wrote the first draft of the manuscript; all authors subsequently took part in the revision process and approved the final copy of the manuscript. H.T., X.N., F.D., V.S., and E.M. provided feedback on the manuscript throughout its development.

Notes

The authors declare no competing financial interest.

ACKNOWLEDGMENTS

The authors thank Evelyn Hu, Marko Lončar, Alexander Ruan, Olivia Mello, and Michaël Lobet for discussion. The Harvard University team acknowledges support from DARPA under contract URFAO: GR510802 for the numerical simulations described in this paper.